\documentclass[prx,twocolumn,final,showpacs,showkeys,nobibnotes,titlepage,twoside,10pt]{revtex4-1} 
\usepackage{amsmath}
\usepackage{graphicx}
\usepackage{amsfonts}
\usepackage{amssymb}
\usepackage{subfigure}
\usepackage{float}
\usepackage{color}

\begin{document}

\title{Local inversion-symmetry breaking controls the boson peak in glasses and crystals}
\author{R. Milkus and A. Zaccone}
\affiliation{Statistical Physics Group, Department of Chemical Engineering and Biotechnology, University of Cambridge, New Museums Site, Pembroke Street, CB2 3RA, Cambridge, U.K. }
\date{\today}

\begin{abstract}

It is well known that amorphous solids display a phonon spectrum where the Debye $\sim \omega^2$ law at low frequency melds into an anomalous excess-mode peak (the boson peak) before entering a quasi-localized regime at higher frequencies dominated by scattering. The microscopic origin of the boson peak has remained elusive despite various attempts to put it in a clear connection with structural disorder at the atomic/molecular level. Using numerical calculations on model systems, we show that the microscopic origin of the boson peak is directly controlled by the local breaking of center-inversion symmetry. In particular, we find that both the boson peak and the nonaffine softening of the material display a strong positive correlation with a new order parameter describing the local inversion symmetry of the lattice. The standard bond-orientational order parameter, instead, is shown to be a poor correlator and cannot explain the boson peak in randomly-cut crystals with perfect bond-orientational order. Our results bring a unifying understanding of the boson peak anomaly for model glasses and defective crystals in terms of a universal local symmetry-breaking principle of the lattice.

\end{abstract}

\maketitle

\section{Introduction}
The phonon spectrum of defect-free crystals is well understood, since the advent of modern solid-state physics in the mid-20th century~\cite{Born1954}. At low frequency and long wavelength, the linear dispersion relation between frequency and momentum results from the breaking of translational symmetry due to the periodic lattice, a manifestation of the Goldstone theorem, and gives rise to the $D(\omega)\sim \omega^2$ Debye law in the density of states (DOS), in 3D. At higher frequencies, phonon propagation through Brillouin-zone boundaries may appear as sharp peaks in $D(\omega)$, known as van Hove singularities~\cite{Maradudin}. 
In the presence of structural disorder, the spectrum of vibrational modes presents very different features which remain poorly understood. The most striking anomaly in glasses is the deviation from the Debye law which manifests itself as the well-documented excess of low-frequency modes visible as a peak in the the normalized DOS $D(\omega)/\omega^{2}$. This effect is widely known as the boson peak anomaly, and is a universal feature in glasses~\cite{Tanaka}, although it has often been observed in crystals as well~\cite{Islam, Bonn, Monaco}.

The Ioffe-Regel crossover~\cite{Ioffe} defines the frequency $\omega_{IR}$ at which the phonon mean-free path becomes equal to its wavelength. Very close to this frequency, is the crossover frequency $\omega^{*}$ 
from ballistic phonons in the linear regime $\omega \sim q$ to quasi-localized modes characterised by diffusive propagation. This crossover is supposed to play an important role for the boson peak in glasses, where local disorder gives rise to scattering at sufficiently small wave-vector $q$, as well as in defective crystals where vacancies and interstitials act as local scattering centres~\cite{Lifshitz,Tanaka,Parshin,Vitelli}. This effect has been attributed, among other mechanisms, to the lowest van Hove singularity in the spectrum of the reference crystalline system, shifted to lower frequencies by disorder-induced level-repulsion effects~\cite{Elliott}. However, no clear or unifying understanding of the role of local structure  has emerged for the boson peak in glasses and defective crystals using standard tools such as e.g. the bond-orientational order~\cite{Nelson,Goodrich}, to characterize the effect of structural disorder. 
Intriguingly, a recent experimental study has shown that the low-$\omega$ peak in the DOS is very similar for the silica glass and the $\alpha$-quartz crystal with matched density~\cite{Monaco}. It was also found that the low-$\omega$ peak in the reduced DOS of $\alpha$-quartz could be accurately reproduced by DFT calculations with just standard phonon physics and interpreted as the lowest van Hove singularity, as shown also in Ref.~\cite{Bosak}.

Here we present numerical results for both model glasses and defective crystals with randomly-cut bonds, and a new conceptual framework to explain the boson peak based on a unifying local symmetry principle. We identify the micro-structural key-factor, which controls the boson peak in both these models of glasses and crystals, with the \textit{local} center-inversion symmetry measured not with respect to the center of the unit cell, but, crucially, with respect to any atom taken as a local center of symmetry.
Given the small number of physical parameters in our model, we discuss each of them carefully in terms of whether they correlate or not with the boson peak.
The local breaking of inversion symmetry appears to be the only microscopic structural signature that correlates with the emergence of the boson peak in both the glass and the defective crystal, without leading to contradictions or paradoxes. 
We show that model glasses and defective crystals having the same average atomic connectivity $Z$, as well as the same density and interatomic interaction, display the same boson peak in spite of having very different values of bond-orientational order. The proposed framework thus naturally provides a unifying framework to explain the boson peak in glasses and defective crystals.

\section{State of the art}
Over the last decades, the vibrational density of states of glasses has been intensely studied from the point of view of theory and simulations, and it is impossible to give a full account of all models. Here we limit ourselves to some of the models that have been applied to experimental data and which connect to our main interest. Among the early models, Thorpe~\cite{Thorpe1976} used a scalar elastic potential for central-force components of interatomic interaction supplemented with bond-bending terms to achieve a description of the DOS of amorphous silicon.  
Subsequently, a model was proposed which is known as the soft-potential model~\cite{Buchenau}, and is largely based on anharmonicity. Within this framework, the anharmonic part of the soft potential causes a redistribution of local oscillator frequencies with a $\sim \omega^4$ scaling in the DOS which then crosses over into a linear $\sim\omega$ scaling. The crossover gives rise to a boson peak in the DOS. 

Another model developed and extensively studied since the 90's is based on the concept of shifted and smeared van Hove singularity. In crystals, van Hove singularities are discontinuities in the slope of the DOS which occur when the dispersion curve is flat, i.e. when $d\omega/dq=0$ at the Brillouin-zone boundaries. In Ref.~\cite{Schirmacher_VH}, a scalar model of a crystal with disorder in the spring constants was studied and the disorder was gradually increased to the point that the van Hove singularity appeared shifted to much lower frequency and much smeared. Since scalar models cannot distinguish between longitudinal and transverse modes, and bear little resemblance to real materials, this model was refined for a vector model by Taraskin et al.~\cite{Elliott}. Here it became clear that it is the transverse van Hove singularity which gets shifted to low frequency thus producing a close resemblance with the boson peak observed experimentally in glasses. Subsequent calculations by Zorn~\cite{Zorn} suggest that the eigenvalues corresponding to the lowered transverse van Hove singularity are presumably those which contribute the most to the boson peak. 
In these models, it is clear that the local atomic packing and structure play a major role in transforming the eigenvalue statistics and distribution of a perfect crystal into the one typical of glasses and defective crystals. 

One of the most popular models of the boson peak was proposed by Schirmacher~\cite{Schirmacher}, and is based on the concept of heterogeneous elasticity. The starting point is the assumption that the shear modulus is a spatially varying quantity due to structural disorder. The microscopic details of atomic arrangements are not specified and the model is entirely a macroscopic one. Yet, the model is very powerful because the assumption of Gaussian quenched disorder allows one to employ the standard tools of statistical field theory to arrive at elegant expressions for both longitudinal and transverse elastic Green functions that are used to calculate the DOS. The assumption of Gaussian disorder has been later generalized to non-Gaussian disorder within the scheme of coherent potential approximation~\cite{Schirmacher2}. Overall, this model provides a picture of the boson peak as a consequence of the crossover from phonon physics (at very large wavelengths) into random-matrix physics at length-scales where the effect of disorder becomes important. 

More recently, important studies, both experimental and computational, have focused on the comparison between crystal polymorphs and glassy phases with the same atomic composition, such as e.g. vitreous silica and $\alpha$-quartz, at matched densities~\cite{Monaco}. It was found that the two systems have a very similar reduced DOS in the boson peak region, and that the boson peak of the glass appears as a smoothened version of the van Hove peak in the crystal. The latter peak in the crystal was shown to be as a lowered and smeared van Hove singularity, as clarified by first-principles numerical calculations~\cite{Bosak}. Also the specific heat for the two systems was found to be identical.

While most theoretical models are limited in their ability to describe the microscopic disorder, numerical simulations have been used in an attempt to identify the microscopic signature of structural disorder which is directly responsible for the boson peak. The most common parameter used to this aim is the bond-orientational order parameter~\cite{Nelson,Goodrich}, which quantifies the spread in the angular orientations of the bonds connecting the atoms in the lattice.
Our main contribution below is to show that in model systems, which share many features of real systems in terms of DOS and elastic properties, the bond-orientational order parameter does not correlate with the boson peak. Instead, the breaking of local inversion symmetry displays a much stronger correlation and proves to be a good candidate for a microscopic structural signature of disorder which could be used as a universal order parameter to link the low-frequency non-Debye peak in the DOS to the underlying atomic structure. This microscopic interpretation is proved here for random network models of glasses and for defective crystals. It remains to see in future studies if this concept can prove useful also for defect-free non-centrosymmetric crystals such as $\alpha$-quartz~\cite{Bosak}. 

Finally, the boson peak in glasses has been found to be dominated by transverse modes and to correlate strongly with softening of the shear modulus~\cite{Tanaka}. We will show below that both the boson peak and the softnening of the shear modulus can be linked to the phenomenon of nonaffine displacements which is caused by the breaking of local inversion symmetry.

\section{Simulation models}
In our simulations, we use a random network created by first randomly placing $N = 4000$ soft spheres in a box and letting them interact via a truncated Lennard-Jones (LJ) potential $V(r) =  (1/r^{12}-2/r^{6}+0.031)\Theta(2 - r)$. The system is brought to a metastable lower energy state by a Monte Carlo energy-relaxation algorithm~\cite{Binder}. Bonds are formed only between nearest neighbours and the bond length is distributed around the mean value $R_0 = 0.94$. The volume of the box is chosen such to create a dense network with an average coordination number $Z=9$ which is almost delta-distributed. The fact that the coordination is the same for all atoms implies the absence of regions which are locally under-coordinated or over-coordinated (with respect to the average $Z$), hence the local rigidity is uniform throughout the sample~\cite{Thorpe}. 
To simulate systems with lower $Z$, we randomly cut bonds from the initial configuration, while keeping a narrow distribution of $Z$. We studied systems with coordination numbers from $Z = 9$ down to $Z = 6$. The density is kept at a constant value $ N/V = 1.467$ and we implemented periodic boundary conditions to avoid surface effects. To reduce noise we calculate our results for ten independent realizations, over which we then take averages. 

\begin{figure}
\centering
\includegraphics[width=0.84\columnwidth]{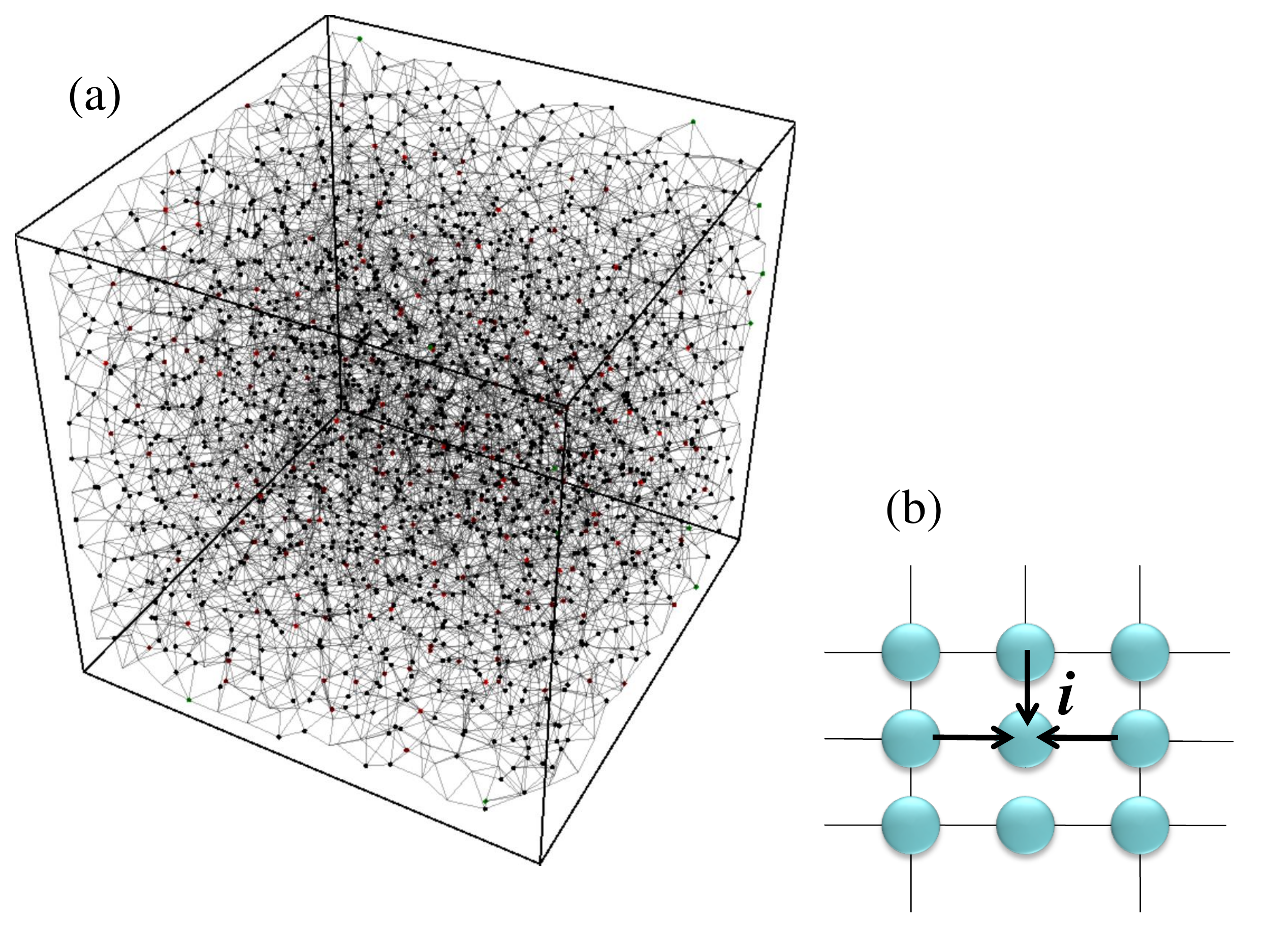}
\caption{(Color online)
(a) One realization of our random network for $Z=6$. The dots represent atoms which are connected by harmonic springs. Local inversion-symmetry is evidently broken by the randomness and lack of correlation in bond orientations. (b) Schematic 2D picture of a regular lattice where locally the removal of a bond breaks the inversion-symmetry on atom $i$. The main consequence of inversion-symmetry breaking induced by the cutting of the bond is the imbalance of NN forces (arrows) acting on atom $i$ when it reaches its affine position under strain. The net force acting on atom $i$ in its affine position has to be released through an additional \textit{nonaffine} displacement. The atom $i$ also acts as a scattering and quasi-localization center~\cite{Lifshitz} for incoming vibrational excitations. 
}
\label{fig:rn}
\end{figure}

In the final step to create our model glass, we then used the so obtained configurations to generate harmonic random-network (RN) model glasses, where the Lennard-Jones interactions between nearest neighbours are all replaced by harmonic springs with pair potential 
$V(r)=(\kappa/2)(r-R_0)^{2}$, with spring constant $\kappa=1$. A sample realization of the model RN glass is shown in Fig.1 for the marginally stable (isostatic) network $Z=6$. We also generated harmonic FCC crystals with the same density and spring constant as the RN glass, and randomly cut bonds to vary $Z$ and to induce the breaking of inversion-symmetry.
This procedure allows us to use all the tools of lattice dynamics and nonaffine linear response theory to analyse the DOS and the shear modulus. The lattice dynamics is governed by the equation of motion for the displacement $\underline{u}_{i}$ of atom $i$, 
$\ddot{\underline{u}}_{i}=-\kappa\sum_{j}\underline{n}_{ij}\cdot(\underline{u}_{i}-\underline{u}_{j})$, with oscillating solutions $\underline{u}_{i}(\underline{r},t)=\underline{u}_{i}(\underline{r})\exp{(i\omega t)}$, leading to $\omega^{2}\underline{u}_{i}=\kappa\sum_{j}\underline{n}_{ij}\cdot(\underline{u}_{i}-\underline{u}_{j})$. Here $\underline{n}_{ij}$ denotes the unit vector pointing from atom $i$ to atom $j$. 
Using the latter relation, the time-independent part of the displacement is related to the dynamical (Hessian) matrix~\cite{Ashcroft} $\underline{\underline{H}}_{ij}= (\partial^2 U/ \partial r_i^{\alpha} \partial r_j^{\beta} )_{\gamma \to 0}$, where $\alpha, \beta=x,y,z$ and its eigenvalues $\lambda$, via $\omega^{2}\underline{u}_{i}=\underline{\underline{H}}_{ij}\underline{u}_{j}=\lambda\underline{u}_{i}$. Hence, $\lambda=\omega^{2}$, upon recalling that the atomic mass is $m=1$. In this way, the phonon density of states $D(\omega)$ is obtained from the diagonalization of the Hessian matrix, from which one obtains the set of eigenvalues $\lambda$. Different eigenvalues are obtained from different realizations of the same sample, and averaging over the realizations leads to the distribution $\rho(\lambda)d\lambda=D(\omega)d\omega$. The DOS is thus calculated for different values of connectivity $Z$, for both the model RN glasses and the FCC crystals with randomly-cut bonds, which allows us to vary $Z$ by keeping the density constant. Also, bonds are severed to always keep a very narrow distribution of $Z$ in all the samples, which ensures that spatial fluctuations of the affine part of the elastic constants is negligible. 

\section{Nonaffine lattice dynamics}
The Hessian matrix is also a key quantity to evaluate the \textit{nonaffine} elastic response of disordered solids.  
The latter is closely connected with the local inversion symmetry of the lattice~\cite{Lemaitre}. In glasses, when applying shear stress to the solid, the atoms tend to reach a new position (\textit{affine} position) proportional to the applied shear strain $\gamma$. In the affine position, the forces transmitted to any atom $i$ by its nearest neighbours (NN) cancel each other out only if atom $i$ is a local center of symmetry. If the atom is not a center of symmetry for the NN bonds, as schematically depicted in Fig.1(b), the NN forces (arrows in Fig.1b) cannot cancel each other out and a net force acting on atom $i$ in the affine position has to be released via an additional \textit{nonaffine} displacement. This is always true for glasses (Fig.1a), but also for crystal lattices with defects or with randomly-cut bonds (Fig.1b), and also for intrinsically non-centrosymmetric crystals like e.g. quartz~\cite{Monaco, Cady}. In the latter systems, however, the covalent character of interatomic bonding, with its non-central component of interaction, makes the applicability of analytical theories of nonaffine lattice dynamics not yet established due to the difficulty of analytically evaluating Hessian matrices with non-central interactions. 
In the harmonic approximation, the total NN force acting on $i$ under a strain $\gamma$ can be expressed as $\underline{f}_i=\underline{\Xi}_{i}\gamma$, where~ $\underline{\Xi}_{i}=-\kappa R_{0}\sum_{j}\underline{n}_{ij}n_{ij}^{x}n_{ij}^{y}$ (see Ref. \cite{Lemaitre}). The vector $\underline{\Xi}_{i}$ plays a very important role because it encodes the local inversion-symmetry of the lattice. As one can easily verify, $\underline{\Xi}_{i}=0$ if atom $i$ is a local center of symmetry, while $\underline{\Xi}_{i}\neq0$ if the lattice does not have local inversion-symmetry. 

Hence, in non-centrosymmetric and disordered lattices, a net total force $\underline{f}_i=\underline{\Xi}_{i}\gamma \neq 0$ acts on any atom $i$ in its affine position. 
Under the action of this force, the atoms have to perform an additional nonaffine displacement into their final \textit{nonaffine} equilibrium positions, which is an internal work contributing negatively to the free energy of deformation, $F(\gamma) = F_{A}(\gamma) - F_{NA}(\gamma)$. The first term, $F_A$, is the contribution from the affine displacements, which is the sum of all the bond-stretching energies, and can be calculated based on the Born-Huang lattice dynamics. The second term, $- F_{NA}$, contains the reduction of the elastic free energy due to the nonaffine relaxation of the system caused by the lack of local inversion-symmetry. 
Recalling that the shear modulus is given by $G=\partial^{2}F/\partial\gamma^{2}$, the local inversion-symmetry breaking thus causes the shear modulus of disordered solids to be lower compared to defect-free centrosymmetric crystals, and as shown in several previous works~\cite{Lemaitre,Zaccone2011,Yoshino}: 
\begin{equation}
G = G_{A} - G_{NA} = G_{A} - \Xi_i^{\alpha} \; (H_{ij}^{\alpha \beta})^{-1}  \Xi_j^{\beta}. 
\end{equation}

Here the second, nonaffine (negative) contribution to the shear modulus $G$ is identically zero only for defect-free centrosymmetric crystal lattices. 
Next we shall use this formalism to evaluate the shear modulus for our model glasses and randomly-cut FCC crystals as a function of the atomic connectivity $Z$. The results are shown in Fig.2.

\begin{figure}
	\centering
	\subfigure
	{\includegraphics[width=0.49\columnwidth]{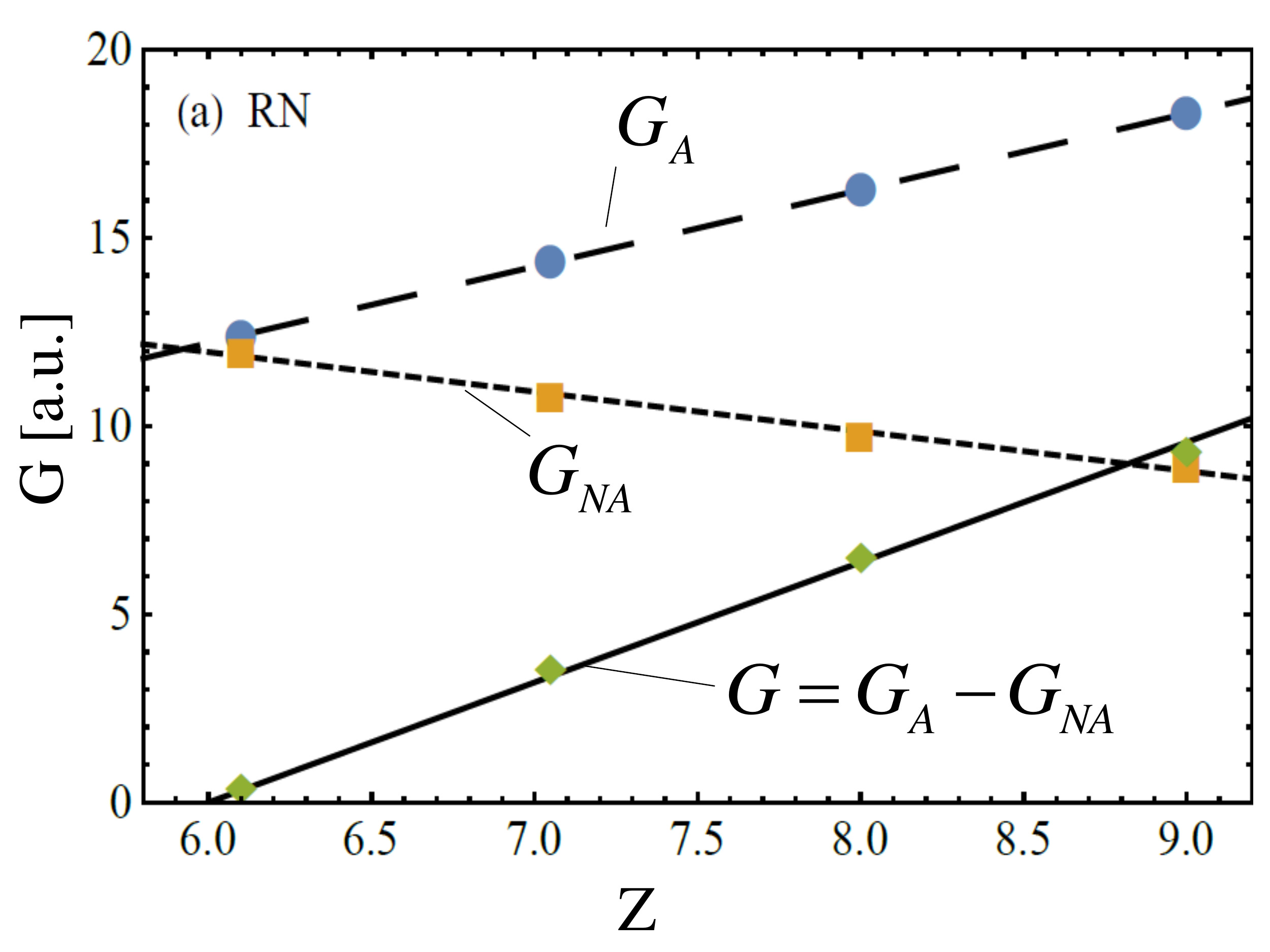}}
    {\includegraphics[width=0.49\columnwidth]{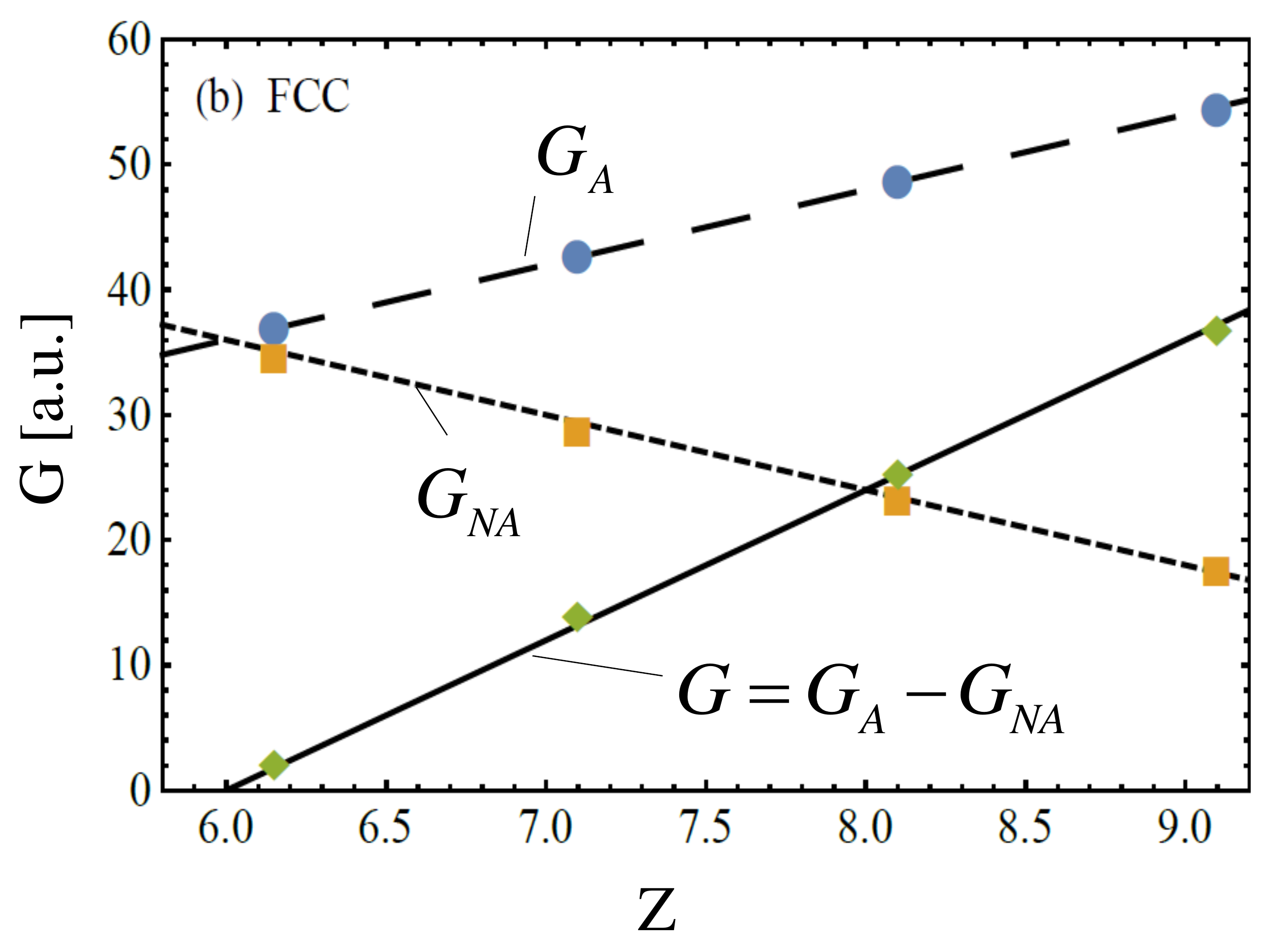}}
	\caption{(Color online) Shear modulus as a function of connectivity. (a) Shear modulus for the RN model glass. (b) Shear modulus of the FCC crystal with randomly-cut bonds.}
	\label{fig:G}
\end{figure}


\section{Analysis of shear elasticity: random network versus defective FCC}
Both our model systems follow the well known $G\sim(Z-6)$ scaling with respect to the isostatic point $Z=6$ found in many previous works. While it is well established~\cite{Lemaitre,Zaccone2011} that $G_{A}\sim Z$, we note here, importantly, that the nonaffine contribution $G_{NA}$ also depends on $Z$, and, in particular, it decreases with increasing $Z$.
In fact, while the affine contribution $G_{A}$ for the glass is in nearly exact quantitative agreement with analytical mean-field predictions for random isotropic networks~\cite{Zaccone2011}, the nonaffine contribution decreases linearly upon increasing $Z$, which deviates from the mean-field theory~\cite{Zaccone2011}. 
From Fig.2 we find that the following law is obeyed: $G_{NA}=a-b(Z-6)$ for both the RN glass and the defective FCC crystal, where $a$ and $b$ are numerical coefficients. For the FCC crystal the nonaffinity vanishes in the limit of the perfect crystal with $Z=12$, and thus $a=6b$. 
Overall, the nonaffinity decreases with increasing $Z$ in qualitatively the same way for both RN lattice and defective crystal, which suggests a common microscopic structural origin for this behaviour, as discussed below. 

\section{Vibrational density of states: random network and defective FCC}
We shall now consider the density of states of both RN glass and defective FCC crystal, for the same conditions investigated for the shear modulus above. The results are shown in Fig.3. At large $Z$-values we observe that, at the lowest frequencies, the parabolic Debye law $D(\omega)\sim \omega^2$ is visible, for both glass and crystal. The only difference in the spectrum is at higher frequencies where two peaks emerge in the FCC spectrum which are reminiscent of the typical peaks in the phonon spectrum of perfect FCC crystals~\cite{Maradudin}; the latter spectrum is eventually recovered at $Z=12$, which we checked. At lower $Z$, where breaking of local inversion symmetry becomes important, the Debye regime shrinks and the boson peak becomes more prominent. Both spectra are quite similar to those of harmonic, stress-free random packings~\cite{Vitelli2}. We also verified that the boson peak frequency scales with connectivity as $\omega_{BP}\sim (Z-6)$. 

The latter scaling can be explained in terms of the crossover between the elastic-continuum regime (ballistic phonon propagation) and the quasi-localized random-matrix-dominated regime (diffusive-like propagation), as suggested in Ref.\cite{Parshin}.

\begin{figure}
\centering
\includegraphics[width=0.98\columnwidth]{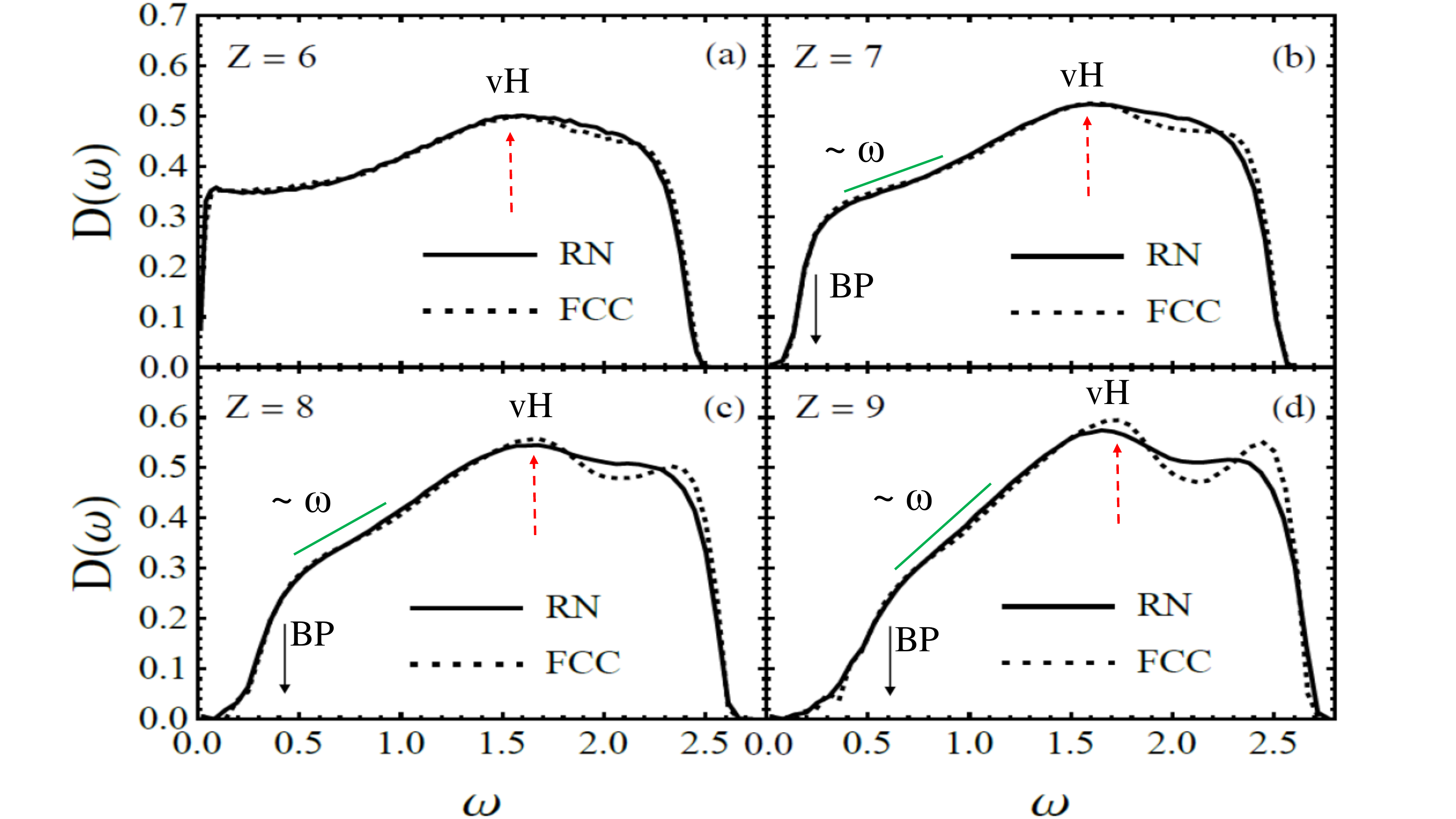}
\caption{(Color online)
Vibrational density of states $D(\omega)$ calculated for the RN glass (solid line) and for the randomly-cut FCC crystal (dotted line), for $4$ different values of atomic connectivity $Z=6,7,8,9$. The solid arrow indicates the approximate position of the boson peak frequency, $\omega_{BP}$, while the dashed arrow indicates the position of the lowest van Hove peak, $\omega_{VH}$. For $Z=6$, $\omega_{BP}\approx0$. The low-energy part of the spectrum, including the boson peak, appears practically identical for the RN glass and for the randomly-cut FCC crystal. 
}
\label{fig:rn}
\end{figure}

\begin{figure}
\centering
\includegraphics[width=0.90\columnwidth]{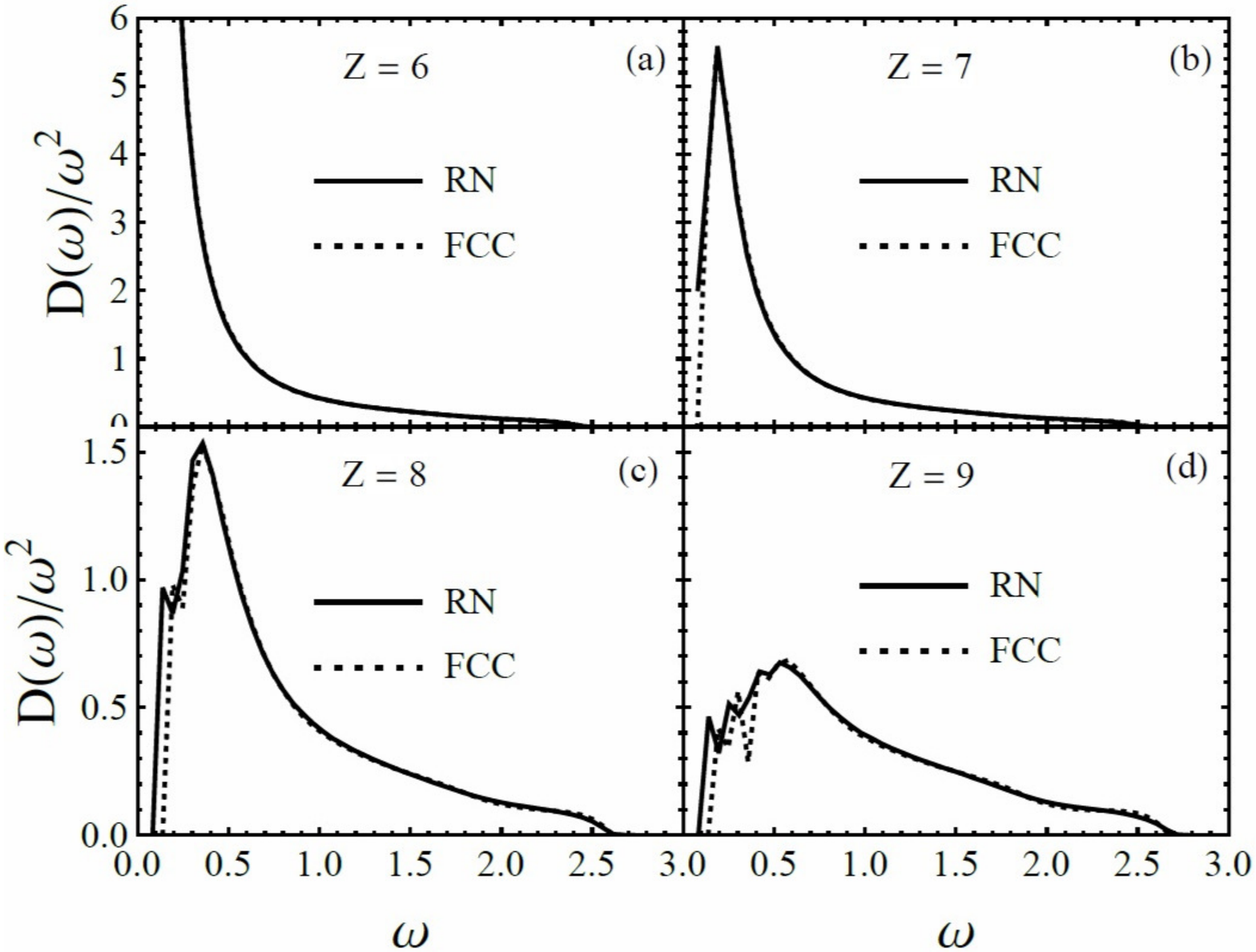}
\caption{(Color online)
Vibrational density of states normalized by the Debye behaviour, $D(\omega)/\omega^{2}$, for the same data of Fig.3.
}
\label{fig:rn}
\end{figure}

\begin{figure}
\centering
\includegraphics[width=0.90\columnwidth]{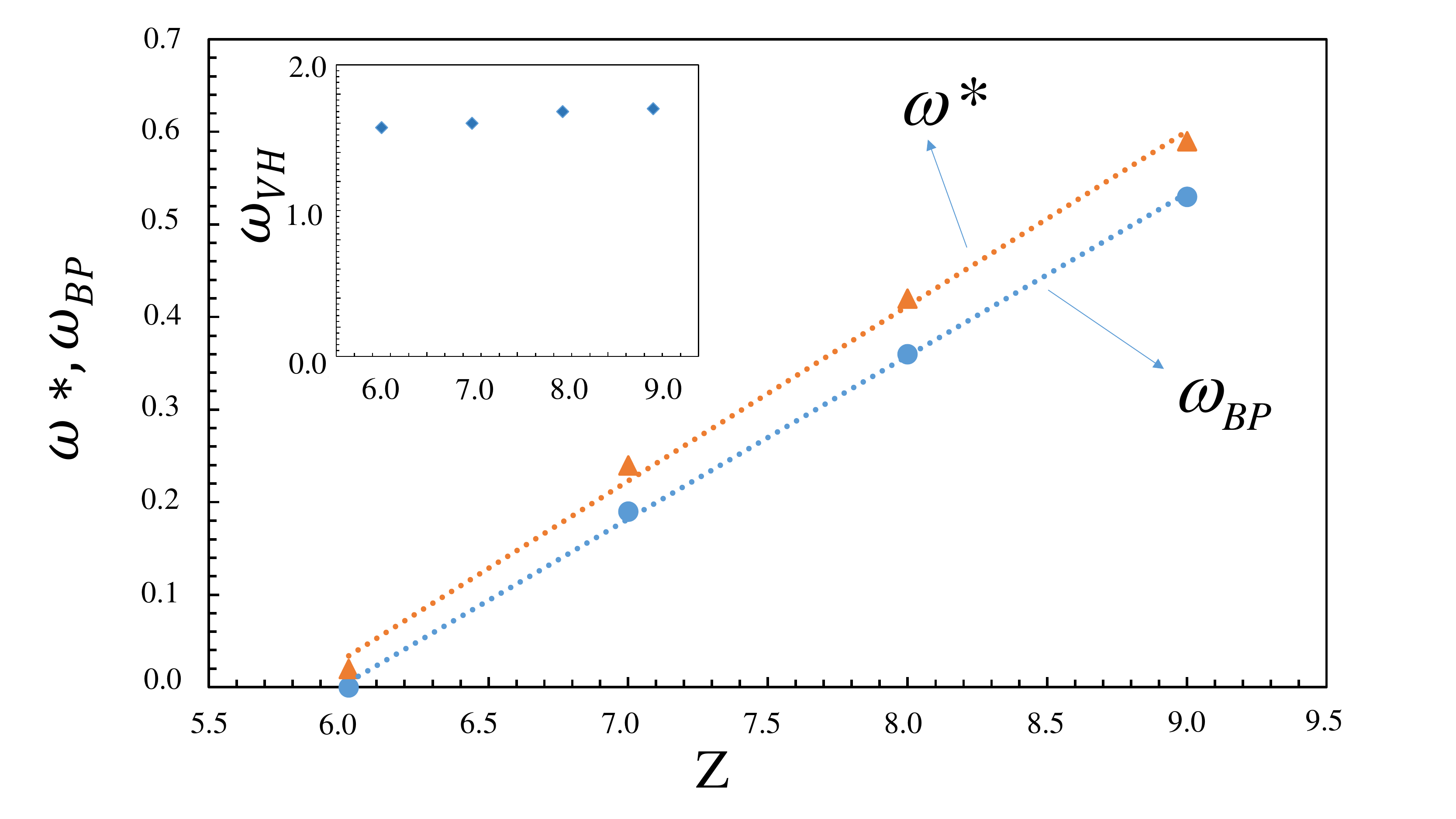}
\caption{(Color online)
Crossover frequency $\omega^{*}$ (which marks the start of the random-matrix regime linear in $\omega$ in the DOS), and the boson peak frequency $\omega_{BP}$, both plotted as a function of the average connectivity $Z$. The data points refer to both random network and defective FCC systems, since both systems have exactly the same values of $\omega^{*}$ and $\omega_{BP}$. The dashed lines are guides to the eye. 
}
\label{fig:rn}
\end{figure}

The most striking fact, in Fig.3 and Fig.4, is that the low-frequency part of the spectrum, including the boson peak, is practically identical for the RN glass and for the randomly-cut FCC crystal. This is an important observation which calls for a mechanistic explanation. 
The structural origin of the boson peak in our system cannot be traced back to spatial fluctuations of the local elastic modulus, because the $Z$ distribution is very narrow by construction, hence the connectivity and the shear modulus are spatially homogeneous. Furthermore, anharmonic effects, also invoked in the past to explain the boson peak~\cite{Buchenau}, obviously play no role because atoms, in our simulations, are connected by strictly harmonic springs. One is then tempted to look for an explanation based on microstructure.
What is really puzzling, however, is that the glass and the randomly-cut FCC crystal display the same boson peak and low-frequency spectrum, in spite of having widely different microscopic structure and disorder. In the RN lattice, NN bonds can have any orientation and the NN unit vector orientation is distributed nearly at random (isotropically) in the solid angle (apart from some weak correlations due to the self-organization of the network). In the randomly-cut FCC crystal, instead, the NN bonds, basically with no exceptions, can have very few orientations only, which are dictated by the crystallographic structure. 

\section{Analysis in terms of the bond-orientational order parameter}
This important microstructural difference between the glass and the randomly-cut crystal becomes evident upon quantifying the bond-orientational order in the two systems. To this aim, we employ the standard bond-orientational order parameter $F_6$, which has been used many times on glasses and defective crystals~\cite{Nelson,Frenkel,Goodrich}. For each pair of NN atoms $i$ and $j$, one first defines the correlator of NN orientations, $S_{6}(i,j)=\frac{\sum_{m=-6}^{6}q_{6m}(i)q_{6m}^{*}(j)}{\mid\sum_{m=-6}^{6}q_{6m}(i)\mid\mid\sum_{m=-6}^{6}q_{6m}(j)\mid}$, where $q_{lm}(i)$ is the usual definition of the local bond-orientational order parameter in terms of spherical harmonics~\cite{Nelson}. 
One then defines the local bond-orientational order parameter as $f_{6}(i)=\frac{1}{Z(i)}\sum_{j}\Theta[S_{6}(i,j)-S_{6}^{0}]$, where $S_{6}^{0}$ is a threshold equal to $0.7$, as discussed in~\cite{Frenkel}, while 
$Z(i)$ is the connectivity of atom $i$ and $\Theta$ the Heaviside function. We finally average $f_{6}(i)$ over all atoms in the system to obtain $F_{6}$. The latter parameter measures the degree of correlation among bond orientations, or in simple words, how many bonds are aligned along the same directions.
Hence, $F_{6}$ has its largest value and is equal to 1 for crystal lattices where all bonds are aligned along the crystallographic orientations.

We thus find $F_6\approx1$ for our randomly-cut FCC crystal under all conditions, as shown in Fig.4. This was expected from the fact that practically all surviving (non-severed) bonds in our randomly-cut crystal are perfectly aligned with the crystallographic directions, and thus have a very high degree of correlation reflected in the $F_6$ being close to 1. This is different from other defective crystals like those studied in~\cite{Goodrich}, where bond-orientational disorder is important because e.g. interstitial atoms introduce bond-orientations which differ from those prescribed by the crystal lattice. The fact that some bonds are not oriented along the crystallographic axes leads, in that case, to $F_6$ values significantly below 1. We also calculated $F_6$ for our model RN glass, and in this case we find a much smaller value, about $0.3$, consistent with the large degree of bond-orientational disorder in our RN glass. We thus face the question of why such widely different degrees of bond-orientational order, for glass and crystal, can coexist with the same boson peak. 

In effect, it appears that the microstructural mechanisms proposed in the past to explain the boson peak, cannot be responsible for the boson peak in our randomly-cut FCC crystal. 
We have already showed above that the key mechanism which controls the softening of the shear modulus in disordered solids is the local inversion-symmetry breaking which is active in both the glass and the randomly-cut crystal, see Fig.1. 
Within the nonaffine response formalism used in our analysis, the nonaffine part of the modulus $G_{NA}$ is also closely related to the density of states $D(\omega)$, and hence to the boson peak, via Eq.(34) of Ref.~\cite{Lemaitre}.
This fact strongly supports the concept we propose here that the local inversion-symmetry breaking is directly related to the emergence of the boson peak. In order to confirm this hypothesis, we shall now quantify the degree of inversion-symmetry breaking in the two systems.

Bond-orientational disorder is known to play a role in the glassy behaviour of orientationally disordered organic crystals~\cite{Bonjour,Ramos}. It is important to notice, however, that, in those systems, orientational disorder is also coupled to anisotropy of the organic molecules. The combination of local orientational disorder and molecular anisotropy leads to breaking the local inversion symmetry, such that the same mechanism of force-imbalance due to local inversion-symmetry in the affine positions (Fig.1) and the ensuing nonaffine softening must be important in those systems as well and should be analysed quantitatively in future work. We thus believe that, also in orientationally disordered crystals, inversion-symmetry breaking, resulting from bond-orientational disorder coupled to molecular anisotropy, can be identified as the source of soft modes and boson peak behaviours.  

\begin{figure}
\centering
\includegraphics[width=0.79\columnwidth]{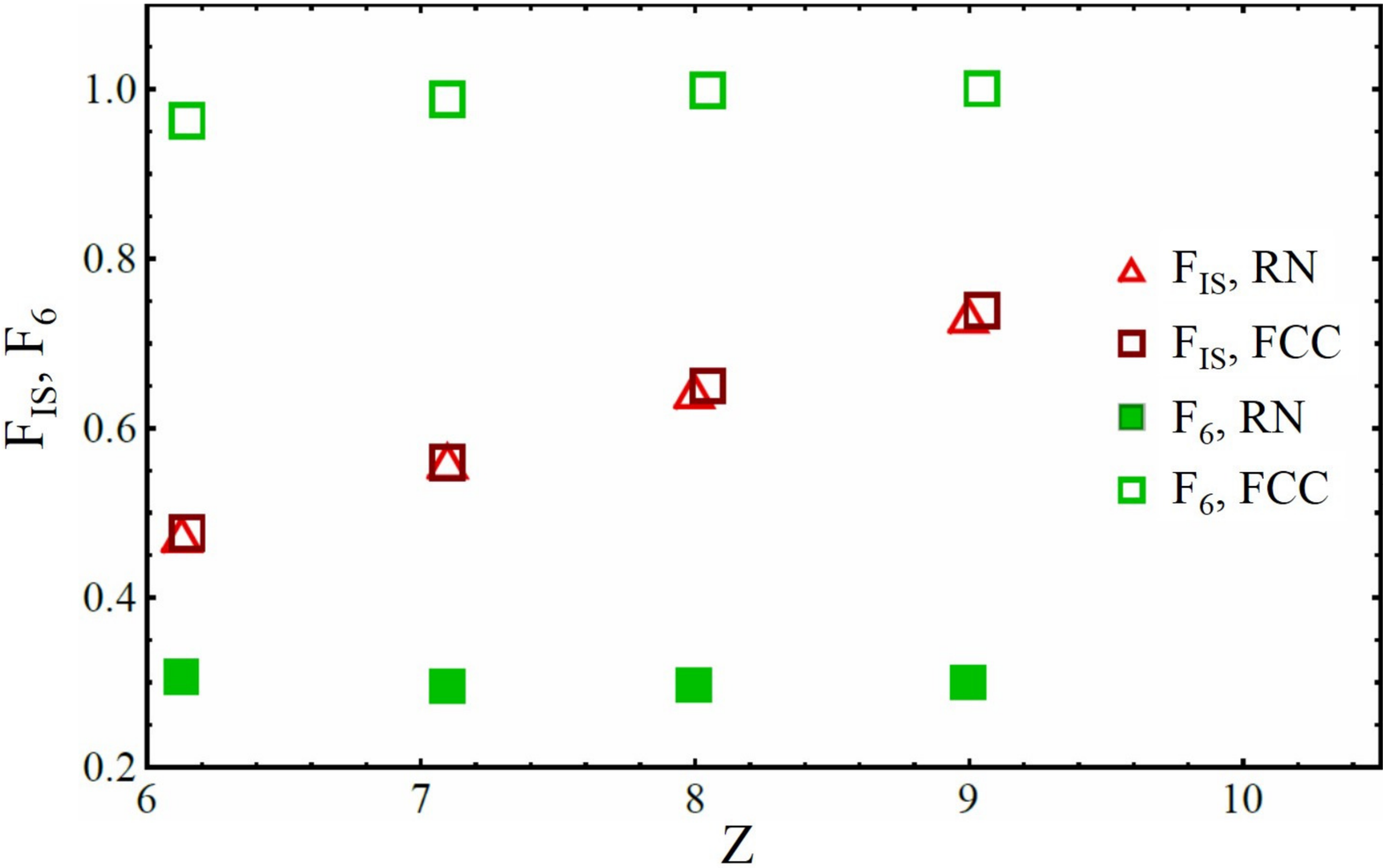}
\caption{(Color online)
Comparison between the inversion-symmetry order parameter ($F_{IS}$) and the standard bond-orientational order parameter ($F_6$) for the RN glasses and for the randomly-cut FCC crystals as a function of connectivity $Z$. There is a perfect collapse of $F_{IS}$ for the two systems onto a master curve as a function of $Z$. 
}
\label{fig:rn}
\end{figure}

\section{A new order parameter for the boson peak based on local inversion-symmetry}
To this aim we propose a new order parameter which, unlike the standard $F_6$, is sensitive to the degree of local inversion-symmetry breaking of the lattice and we shall test how it correlates with both the shear modulus and the boson peak. A good starting point is the absolute value of the sum of all nearest-neighbour force vectors (squared) in the affine configuration (affine force vectors) $|\underline{\Xi}|^{2}$, which is identically zero for perfect centrosymmetric crystal lattices and has its largest values for lattices where the local inversion symmetry is completely absent. To measure the degree of symmetry breaking independent of the direction of deformation, we additionally sum over all possible Carthesian coordinate pairs $|\underline{\Xi}|^{2} \equiv \sum_{\alpha,\beta \in \{x,y,z\}} |\underline{\Xi}_{\alpha \beta}|^{2}$. The order parameter for local inversion-symmetry is thus defined as 
$F_{IS}=1-\frac{\sum_{\alpha,\beta \in \{x,y,z\}} |\underline{\Xi}_{\alpha \beta}|^{2}}{\sum_{\alpha,\beta \in \{x,y,z\}} |\underline{\Xi}_{\alpha \beta}|^{2}_{ISB}}$, where 
$\mid\underline{\Xi}_{\alpha \beta}\mid^{2}_{ISB}$ indicates the limit in which inversion symmetry is completely broken and there cannot be any correlations whatsoever between bond orientations. For the latter case, we found
$|\underline{\Xi}_{\alpha \beta}|^2_{ISB} \,=\, \kappa^2 R_0^2 \sum_{i j} \left(n_{i j}^\alpha n_{i j}^\beta \right)^2$, a result derived in Appendix C. Assuming that each lattice site has the same coordination number $Z$, we can simplify the denominator to $\sum_{\alpha,\beta \in \{x,y,z\}} |\underline{\Xi}_{\alpha \beta}|^{2}_{ISB} = \kappa^2 R_0^2 N Z$. 
Hence, $F_{IS}=1$ for any \textit{perfect} centrosymmetric lattice, while $F_{IS}=0$ for the limiting configuration at which the local breaking of inversion-symmetry is maximum. 

The new $F_{IS}$ order parameter has a strong correlation with the nonaffine part of the shear modulus, reflected in the empirical relation $G_{NA}\propto\langle\mid\underline{\Xi}_{i}\mid^{2}\rangle/Z\propto(Z_{0}-Z)/Z_{0}$, with $Z_{0}=12$ for the FCC case, which we obtain from the simulations.
Importantly, the values of the IS order parameter for both FCC and random network appear to be basically the same in Fig.5. This crucial observation lends further support to the conclusion that the boson peak is controlled by inversion symmetry and this is the only possible explanation to the fact that the boson peak is exactly the same for FCC and RN lattices.

The order parameter for local inversion symmetry, $F_{IS}$, is plotted in Fig.6, in comparison with the standard bond-orientational order parameter $F_6$.
It is seen that $F_{IS}$ displays the linear trend with $Z$ which correlates well with both the $Z$-dependence of boson peak frequency, and with the nonaffine shear softening, also linear in $Z$. Further, $F_{IS}$ displays very similar values, for both the glass and the crystal, at any given $Z$, which also appears consistent with the boson peaks being the same for both systems in Fig.3 and Fig.4. 
No such correlation is displayed by $F_6$, which remains always constant with $Z$, and has widely different values for the RN glass and the defective crystal, in Fig.6.

\section{Analysis of physical parameters in our model and their relation, or lack thereof, to the boson peak}
In our model system there are very few physical parameters at play. It is therefore possible to check for each of them separately whether they correlate with the boson peak or not. 
\subsection{Density $N/V$}
This parameter cannot control the boson peak. The density, defined as $N/V$, i.e. the total number of atoms per unit volume, is constant in all our simulations, and is constant with $Z$. The connectivity $Z$ is decreased not because we decrease the number of atoms per unit volume, which remains always the same, but because we cut bonds connecting atoms. This fact proves that the density plays no role and does not affect the boson peak. 
\subsection{Force constant $\kappa$}
The value of force constant $\kappa$ for our harmonic springs is never varied in our simulations. Hence, this parameter cannot control the boson peak. 
\subsection{Lowest van Hove singularity}
This parameter does not correlate with the boson peak frequency either. We have added a dashed arrow in Fig.3 to mark the position of the lowest van Hove peak in the DOS. At the highest $Z$ considered, both van Hove peaks in our defective FCC lattice are clearly visible, and they occur at the same frequencies expected for the perfect FCC lattice (as one can easily check). Upon decreasing $Z$, the boson peak develops, and its frequency shifts to lower values. The van Hove peak, instead, remains at the same frequency $\omega_{VH}$, which is a much higher frequency (about three times higher) than the boson peak frequency. As is shown in the inset of Fig. 5, $\omega_{VH}$ remains constant upon decreasing $Z$ in our defective FCC crystal, hence it does not correlate with the boson peak. 
Another important observation which confirms that our boson peak is not due to the van Hove singularity comes from the scaling of the DOS at frequencies just above the boson peak. The scaling is $D(\omega)= A\omega+B$ in this regime, which means that the eigenvalue distribution scales as $\rho(\lambda)= (A/2) +(B/2) \lambda^{-1/2}$, upon recalling the definition $\omega=\lambda^{2}$. Below the boson peak, and for $Z>6$, the behaviour is instead $D(\omega)\sim \omega^{2}$, i.e. fully consistent with Debye law. Upon approaching $Z=6$, the coefficient $A$ vanishes and we reproduce the random-matrix scaling  $\rho(\lambda)\sim \lambda^{-1/2}$ found analytically in the Marcenko-Pastur distribution of random matrix theory. For $z>6$ the Debye regime arises and alters this scaling. This important observation shows that the boson peak coincides with the crossover from phonon wave propagation to a regime dominated by scattering and by random-matrix eigenvalue statistics. A physical explanation for this may be found in the fact that atoms which are no longer local centers of inversion symmetry act as scattering centers for incoming waves.
The same type of crossover from Debye regime to random-matrix behaviour across the boson peak has been found by Zamponi, Parisi and coworkers for a mean-field model of glasses~\cite{Zamponi}. For the crossover from ballistic phonons to quasi-localized modes see also Ref.~\cite{Parshin}.

A further indication that the boson peak in our systems is of a different nature from the van Hove singularity, comes from the analysis of localization. Upon considering the plot of the participation ratio of vibrational modes $p(\omega)$, reported in in Appendix A below, it is evident that both the random network and the defective FCC have localized or quasi-localized modes at about the same frequency at which the boson peak is observed. In particular, for $Z=7$, both the random network and the defective FCC have values of participation ratio approaching zero near the boson peak. It is clear that such a strong localization of vibrational modes at the boson peak frequency is not compatible with an explanation based on standard phonon physics and the van Hove singularity.

\subsection{Connectivity Z}
There is an obvious correlation between $Z$ and the boson peak (both BP frequency and amplitude), as there is also an evident correlation between Z and the inversion-symmetry order parameter $F_{IS}$. However, we believe that the boson peak depends on $Z$ mainly because the boson peak depends on the local inversion-symmetry, which in turn is controlled by $Z$ due to the bond-cutting method.
The reason why we do not believe that $Z$ can be the ultimate cause of the boson peak is that, if this was true, then we should observe a very strong boson peak also in the simple cubic (SC) lattice which has $Z=6$, exactly. This is plainly impossible. 
In our systems (random network and defective FCC) at $Z=6$ there is no trace left of Debye behaviour (see our Fig. 3 above), whilst there is a very strong boson peak at vanishing frequency, $\omega\rightarrow 0$. Instead, in the nearest-neighbour SC lattice with $Z=6$ there is obviously perfect Debye behaviour and no trace of boson peak; the dispersion relation for the SC lattice can be calculated analytically which gives $\omega=\sqrt{4(\kappa/m)\sin^{2}(q/2)}$, where $q$ is the wave-vector. One should therefore conclude that connectivity alone, or even isostaticity~\cite{Wyart}, cannot explain the boson peak because there is no boson peak in isostatic structures such as the SC lattice with $Z=6$. 
\subsection{Bond-orientational disorder}
As discussed with reference to Fig.6, bond-orientational disorder does not correlate with the boson peak, as in fact it remains constant while the boson peak and its frequency change significantly. 
\subsection{Local inversion-symmetry breaking}
This is the only physical parameter which correlates strongly with the boson peak and does not imply any paradox or contradiction. 

\section{Conclusion}
We have studied two numerical models of disordered solids: a disordered glass with bond-orientational disorder ($F_6\approx0.3$), and an FCC crystal with randomly-cut bonds and perfect bond-orientational order ($F_6\approx1$). 
In spite of the widely different bond-orientational disorder, the two systems exhibit exactly the same boson peak and almost the same nonaffine softening of the shear modulus. In particular, we showed that in both cases the boson peak frequency and the shear modulus display the same scaling with connectivity, which correlates strongly with the degree of local inversion symmetry. 
Since this observation in our system cannot be explained based on other mechanisms invoked in previous models, we arrived at the conclusion that the most likely microscopic origin of both boson peak and nonaffinity resides in the local inversion-symmetry breaking in the lattice, which is very important for both the glass and the randomly-cut crystal. This conclusion is supported by a new order parameter for centrosymmetry which displays a strong correlation with both the boson peak and the nonaffine modulus, for both the glass and the crystal. Within this new framework, the boson peak is caused by the scattering of vibrational modes on atoms which are not local centers of symmetry; such scattering and quasi-localization~\cite{Lifshitz} events become important at nanometric length-scales (frequencies) comparable to the first coordination shells, as shown in previous simulation studies~\cite{Barrat}. 
Finally, our analysis identifies new local structural signatures of soft modes, which cannot be traced back to purely structural quantities (e.g. the structure factor or bond-orientational order parameters), yet they bear an important relation to dynamical heterogeneities because local configurations lacking inversion symmetry would be inherently unstable already under thermal stresses, e.g. in supercooled liquids~\cite{GarrahanPRL}. These local mechanical instabilities could correlate with local regions of dynamical activity. In future work it will be important to ascertain the existence of quantitative correlations between our proposed inversion-symmetry order parameter and the dynamical activity order parameter used to quantify and map dynamical heterogeneity and elastic heterogeneity~\cite{Schirmacher,Mizuno} at the glass transition in relation with soft modes~\cite{Garrahan1,Garrahan2}. 
This proposed shift in paradigm, and the proposed new order parameter, can be used in future studies, with the aid of new theoretical concepts~\cite{Zamponi}, to arrive at a unified understanding of amorphous materials.

\section{Appendix A. Participation ratio of vibrational modes}
For the random network (RN) glass and the randomly-cut defective FCC crystals studied in this work, we also calculated the participation ratio of vibrational modes, in order to determine to which extent the modes are localized or delocalized, as a function of frequency. 
The participation ratio is defined as follows~\cite{Vitelli}:
\begin{equation}
p(\omega)=\left[ N \sum_{i=1}^{N}\mid\underline{e}_{i}\mid^{4}(\omega)\right]^{-1}.
\end{equation}
In this expression, $\underline{e}_{i}$ is the projection of the normalized eigenvector with frequency $\omega$, onto atom $i$, or in other words, the displacement on atom $i$ belonging to the collective vibrational mode $\omega$. 
By construction, $p(\omega)=1$ for ballistic phonons, while it is equal to zero for completely localized modes.
The participation ratio is plotted below for the different values of connectivity $Z$, for both RN glass and FCC crystal.

The qualitative behaviour is very similar to the one observed in simulations of random packings~\cite{Vitelli} and harmonic lattices with spring-constant disorder~\cite{Parshin}. In the low frequency regime, where the linear dispersion relation and the Debye law are valid, the participation ratio is always large and very close to 1, as expected for phonons. Then the participation ratio goes through a minimum corresponding approximately to the boson peak frequency, which is also close to the Ioffe-Regel frequency at which the physics changes from phonons to random-matrix transport. This frequency also corresponds to the wavevector or length scale at which scattering of collective vibrational modes due to local inversion-symmetry breaking becomes important. 

\begin{figure}
\centering
\includegraphics[width=0.84\columnwidth]{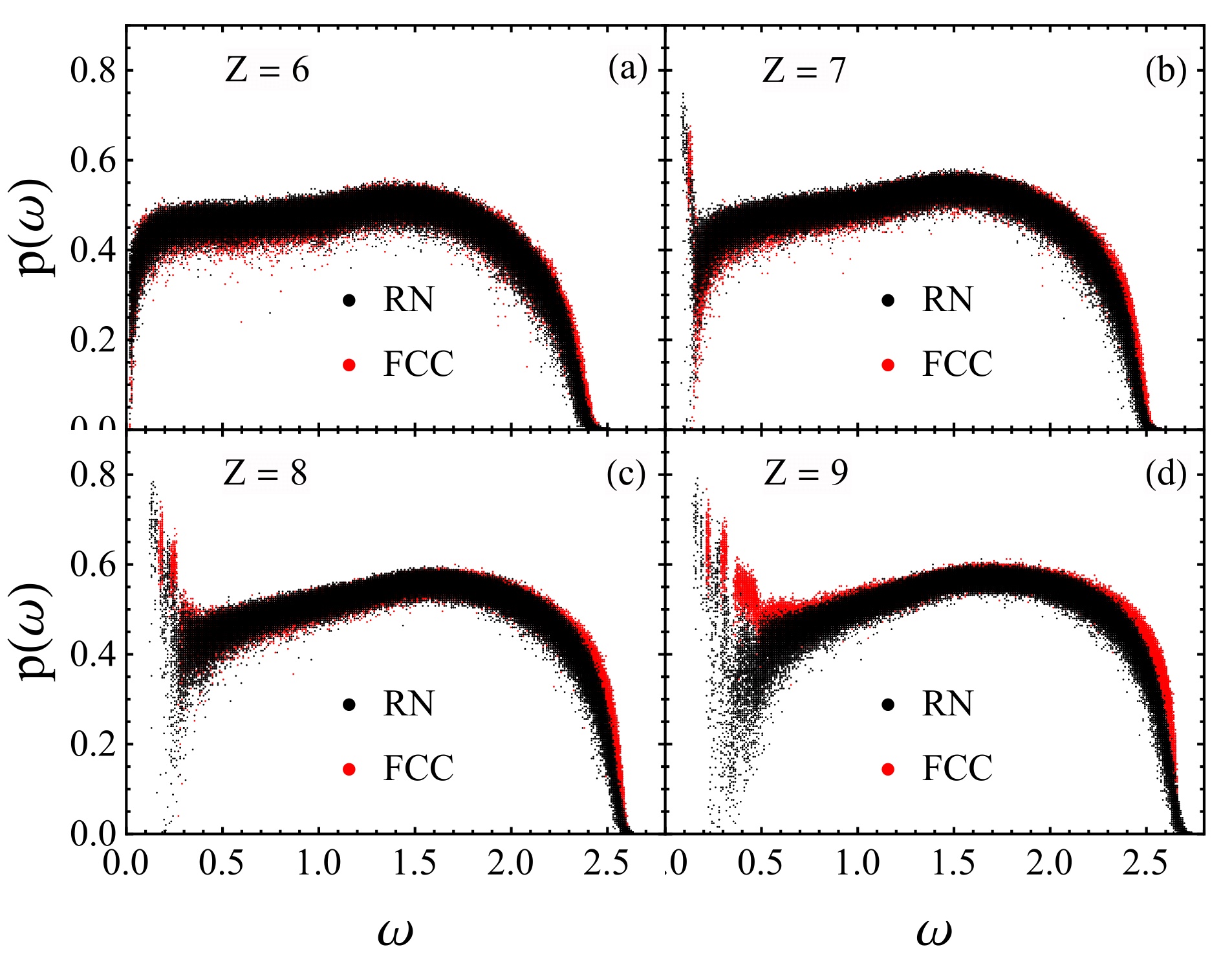}
\caption{
Participation ratios calculated for the RN glass and for the randomly-cut FCC crystal, for different values of average connectivity $Z$. 
}
\label{fig:rn}
\end{figure}

These scattering events cause the modes to become quasi-localized~\cite{Lifshitz}, which is reflected in much lower values of $p(\omega)$. The part of the spectrum just above the boson peak is dominated by randomness, and by an eigenvalue distribution characterized by the scaling $\rho(\lambda)\sim \lambda^{1/2}$, typical of random-matrix models~\cite{Zamponi}.  
Finally at the highest frequencies of the spectrum, close to the Debye frequency, the participation ratio approaches zero for Anderson-localized modes.

%

\section{Appendix B. Order parameter for inversion-symmetry breaking}
We present here a derivation of the analytical expression for the inversion-symmetry order parameter $F_{IS}$ for the case of defective FCC crystals with randomly-depleted bonds. 
We start from a generic defective FCC system with a distribution of bond angles $\theta$ and $\phi$, which define the orientation of a bond unit vector $\underline{n}_{ij}=(\cos\phi\sin\theta, \sin\phi\sin\theta, \cos\theta)$ between two atoms $i$ and $j$. In the framework of the affine force field, for every bond vector 
$\underline{n}_{ij}$ there exists a vector $\underline{n}_{ji} = - \underline{n}_{ij}$ with the same probability $\rho(\theta,\phi)$ in the solid angle. We now write the general expression of the total affine force field 
$|\underline{\Xi}|^2$, as
	\begin{equation}
	|\underline{\Xi}|^2 \,=\, \kappa^2 R_0^2 \sum_{i} \sum_{\alpha }\left( \sum_{j\, nn\, i} {n}_{i j}^\alpha n_{i j}^x n_{i j}^y \right)^2
	\end{equation}
	where $\alpha = x,y,z$ are the Cartesian coordinates. We can carry out those sums and regroup the terms to get
	\begin{widetext}
	\begin{equation}\label{AFexp}
	|\underline{\Xi}|^2 = \kappa^2 R_0^2 \left(\sum_{i j} \left(n_{i j}^x n_{i j}^y \right)^2 +\sum_{i} \sum_{k, l \, nn\, i} (\underline{n}_{i k} \cdot \underline{n}_{i l})(\underline{n}_{i k} \cdot \underline{n}_{i l})^x (\underline{n}_{i k} \cdot \underline{n}_{i l})^y \right).
	\end{equation}
	\end{widetext}
	Now we implement the difference between the most asymmetric configuration where inversion symmetry is completely broken, which we call the ISB, and any other configuration that we want to calculate the order parameter for, such as e.g. a defective FCC crystal. \\
	If there are no constraints whatsoever on the angular correlations between bonds connecting to the same atom $i$ the center of a unit cell, the second term in \eqref{AFexp} is zero. We can explain this by the fact that, as mentioned above, the probability to have any bond vector according to a given angular distribution is equal to the probability to have the negative of this vector (same orientation, opposite direction). In the framework of the scalar product, this means that, for the probability of the quantity in the second right-most sum in Eq.\eqref{AFexp}, the following equality must hold
	\begin{widetext}
	\begin{equation}\label{AFav}
	\begin{gathered}
		\rho((\underline{n}_{i k} \cdot \underline{n}_{i l})(\underline{n}_{i k} \cdot \underline{n}_{i l})^x (\underline{n}_{i k} \cdot \underline{n}_{i l})^y)\,=\,\rho(-(\underline{n}_{i k} \cdot \underline{n}_{i l})(\underline{n}_{i k} \cdot \underline{n}_{i l})^x (\underline{n}_{i k} \cdot \underline{n}_{i l})^y)\\
		\longrightarrow \left\langle 	(\underline{n}_{i k} \cdot \underline{n}_{i l})(\underline{n}_{i k} \cdot \underline{n}_{i l})^x (\underline{n}_{i k} \cdot \underline{n}_{i l})^y \right\rangle \,=\, 0,
	\end{gathered}
	\end{equation}
	\end{widetext}
where the averaging denotes the isotropic angular averaging $\langle ... \rangle = \int ... \frac{1}{4\pi}\sin\theta d\theta d\phi$.
	In a hard sphere system, one has the constraint that $\underline{n}_{i k} \cdot \underline{n}_{i l} < 0.5$, since two bonds both connected to the same atom $i$ cannot have an angle smaller that $\pi/3$ (ultimately due to excluded volume). This constraint shifts the average in \eqref{AFav} from zero to a negative value and lowers the final value of 
	$|\underline{\Xi}|^2$. This is so because the excluded volume correlations raise the average degree of inversion symmetry in the system with respect to a system where the excluded volume constraint on the angles the bonds is absent. In a system where no correlations exist between bond orientations such that the breaking of inversion symmetry is maximum and the local bond orientations are completely asymmetric, the only term which remains in the expression of $|\underline{\Xi}|^2$ is
	\begin{equation}\label{AFiso}
	|\underline{\Xi}|^2_{ISB} \,=\, \kappa^2 R_0^2 \sum_{i j} \left(n_{i j}^x n_{i j}^y \right)^2.
	\end{equation}
	Therefore our order parameter becomes
	\begin{equation}\label{OrderParameter}
	F_{IS}\,=\,1\,-\,\frac{|\underline{\Xi}|^2}{\kappa^2 R_0^2 \sum_{i j} \left(n_{i j}^x n_{i j}^y \right)^2}.
	\end{equation}
	This expression can be easily evaluated numerically for different lattices and provides the correctly normalized limit used in the main article to plot $F_{IS}$ for both the FCC crystal and the RN lattice as a function of $Z$.

\section{Appendix C. Analytical expression for the $F_{IS}$ order parameter for defective FCC crystals}
We will now derive the analytical value for the affine force field of the depleted FCC lattice in order to get an analytical expression for $F_{IS}$. To this aim, we have to calculate $|\underline{\Xi}_{\alpha \beta}|^2 = |{\Xi}_{\alpha \beta}^x|^2 + |{\Xi}_{\alpha \beta}^y|^2 + |{\Xi}_{\alpha \beta}^z|^2$ for the two cases $\alpha =\beta$ and $\alpha \neq \beta$. We start with the general definition of the affine force field on a generic atom $i$:
	\begin{equation}\label{AF}
	\Xi_{\alpha \beta, i}^\gamma = - R_0 \kappa \sum_{j} n_{i j}^\gamma n_{i j}^\alpha n_{i j}^\beta,
	\end{equation} 
where $\alpha, \beta, \gamma$ are Carthesian directions.
	Since no Carthesian direction or plane is in any way special, we can pick one example for each of the two cases. So we explicitly calculate $|\underline{\Xi}_{x x}|^2$ and $|\underline{\Xi}_{x y}|^2$. In the first case the x component of the affine force field is:
	\begin{widetext}
	\begin{equation}\label{xxx}
	|{\Xi}_{x x}^x|^2 = R_0^2 \kappa^2 N \sum_{i = 0}^{8} \sum_{j = 0}^{i} \frac{(2 j - i)^2}{8} \frac{{4 \choose j}{4 \choose i - j}{4 \choose Z - i}}{{12 \choose Z}} = R_0^2 \kappa^2 N \frac{Z (12 - Z)}{132}.
	\end{equation}
	\end{widetext}
	Here $N$ is the number of particles in the system. In the x-component we have 8 allowed bond orientations that can contribute to the affine force field $\underline{\Xi}_{x x}$. Out of a given value of $Z$ bonds in the unit cell,  only $i$ contribute in the x direction. $j$ out of those $i$-contributing bonds give a positive contribution in the sum of \eqref{AF}, thus $(i-j)$ give a negative contribution to the sum. The absolute value of the sum is then $(j - (i - j))$ times the value that each bond contributes, which is $R_0 \kappa/2 \sqrt{2}$. Since we want to calculate the absolute square of the affine force field, we have to consider the square of this value, which gives $R_0^2 \kappa^2/8$. For the y- and z-component we get similar expressions with the difference that now only 4 bonds contribute, in each of these two directions. for example, for the $y$ component we get:
	\begin{widetext}
	\begin{equation}\label{yxx}
	|{\Xi}_{x x}^{y}|^2 = R_0^2 \kappa^2 N \sum_{i = 0}^{4} \sum_{j = 0}^{i} \frac{(2 j - i)^2}{8} \frac{{2 \choose j}{2 \choose i - j}{8 \choose Z - i}}{{12 \choose Z}} = r_0^2 \kappa^2 N \frac{Z (12 - Z)}{264},
	\end{equation}
	\end{widetext}
and we get exactly the same for the $z$ component.
	Now we just sum up the x-,y- and z-component to get
	\begin{equation}\label{xx}
	|\underline{\Xi}_{x x}|^2 = |{\Xi}_{x x}^x|^2 + |{\Xi}_{x x}^y|^2 + |{\Xi}_{x x}^z|^2 = R_0^2 \kappa^2 N \frac{Z (12 - Z)}{66}.
	\end{equation}
	We can use these results to easily calculate $|\underline{\Xi}_{x y}|^2$. The x- and y-components are equal to the y- and z-component of \eqref{xx}, as they correspond to a sum in which two of the indexes $\alpha,\beta,\gamma$ in \eqref{AF} are equal while one is different. This means that we have 4 contributing bonds and can apply Eq.\eqref{yxx}. The z-component is 0, since any product of the three different components of the each unit bond vector vanishes in this system. So we get:
	\begin{widetext}
	\begin{equation}\label{xy}
	|\underline{\Xi}_{x y}|^2 = |{\Xi}_{x y}^x|^2 + |{\Xi}_{x y}^y|^2 + |{\Xi}_{x y}^z|^2 = |{\Xi}_{x x}^y|^2 + |{\Xi}_{x x}^z|^2 + 0 = R_0^2 \kappa^2 N \frac{Z (12 - Z)}{132}
	\end{equation}
	\end{widetext}
	Now we can calculate:
	\begin{widetext}
	\begin{equation}\label{AFtot}
	\sum_{\alpha,\beta = x,y,z}|\underline{\Xi}_{\alpha \beta}|^2 =
	|\underline{\Xi}_{x x}|^2 +|\underline{\Xi}_{x y}|^2 +|\underline{\Xi}_{y x}|^2 +|\underline{\Xi}_{y y}|^2 +|\underline{\Xi}_{y z}|^2 +|\underline{\Xi}_{z y}|^2 +
	|\underline{\Xi}_{z z}|^2 +|\underline{\Xi}_{z x}|^2 +|\underline{\Xi}_{x z}|^2 
	= 	R_0^2 \kappa^2 N \frac{Z (12 - Z)}{11}
	\end{equation}
	\end{widetext}
	If we insert this into Eq.(6) of this appendix upon evaluating the denominator in mean-field approximation, the expression $F_{IS} = 1 - \sum_{\alpha,\beta}|\underline{\Xi}_{\alpha\beta}|^2/R_0^2 \kappa^2 N Z$ leads to the following simple analytical relation:
	\begin{equation}
	F_{IS} = 1 - \frac{\sum_{\alpha,\beta} |\underline{\Xi}_{\alpha\beta}|^2}{R_0^2 \kappa^2 N Z} = 1 - \frac{12 - Z}{11} = \frac{Z - 1}{11}.
	\end{equation}
This situation, where $F_{IS}=0$ and $Z=1$ could be achieved for example in a liquid where most nearest-neighbours are short-lived and highly fluctuating, and only one mechanical bond, on average per atom, is active.

\begin{acknowledgments}
Many useful discussions with E.M. Terentjev, P. Schall, and D. Bonn are gratefully acknowledged.

\end{acknowledgments}


\begin{thebibliography}{99}

\bibitem{Born1954}
M. Born and K. Huang, \emph{Dynamical Theory of Crystal Lattices} (Oxford University Press, 1954).

\bibitem{Maradudin}
G.K. Horton and A. A. Maradudin, \emph{Dynamical Properties of Solids} (North-Holland, Amsterdam, 1974).

\bibitem{Tanaka}
H. Shintani and H. Tanaka, Nat. Mater. \textbf{7}, 870 (2008).

\bibitem{Islam}
D. Kaya, N.L. Green, C.E. Maloney, M.F. Islam,  Science \textbf{329}, 656 (2010).

\bibitem{Bonn}
R. Zargar, J. Russo, P. Schall, H. Tanaka, and D. Bonn, EPL \textbf{108}, 38002 (2014).

\bibitem{Monaco}
A.I. Chumakov, et al., Phys. Rev. Lett. \textbf{112}, 025502 (2014).


\bibitem{Ioffe} A. F. Ioffe and A. R. Regel,Prog. Semicond. \textbf{4}, 237 (1960).

\bibitem{Lifshitz} I. M. Lifshitz, J. Phys. (USSR) \textbf{7}, 215 (1943). 

\bibitem{Parshin}
Y.M. Beltukov, V.I. Kozub, D.A. Parshin, Phys. Rev. B 87, 134203 (2013).

\bibitem{Vitelli}
V. Vitelli, N. Xu, M. Wyart, A.J. Liu, and S.R. Nagel,  Phys. Rev. E \textbf{81}, 021301 (2010).


\bibitem{Elliott}
S.N. Taraskin, Y.L. Loh, G. Natarajan, and S.R. Elliott, Phys. Rev. Lett. \textbf{86}, 1255 (2001).

\bibitem{Nelson}
P.J. Steinhardt, D.R. Nelson, and M. Ronchetti, Phys. Rev. B \textbf{28}, 784 (1983).


\bibitem{Goodrich}
C. P. Goodrich, A.J. Liu, and S.R. Nagel, Nat. Phys. \textbf{10}, 578 (2014).

\bibitem{Bosak}
A. Bosak, M. Krisch, D. Chernyshov, B. Winkler, V. Milman, K. Refson, C. Schulze-Briese, Zeit. Kristall. 227, 84 (2012). 


\bibitem{Thorpe1976}
M.F. Thorpe, in \textit{Physics of Structurally Disordered Solids}, S.S. Mitra Ed. (Plenum Press, Singapore, 1976).


\bibitem{Buchenau} U. Buchenau, Y. M. Galperin, V. L. Gurevich, and H. R.
Schober, Phys. Rev. B 43, 5039 (1991); V. L. Gurevich and H. R. Schober, Phys. Rev. B 57(11), 295
(1998); V. L. Gurevich, D. A. Parshin, and H. R. Schober, Phys. Rev. B 67, 094203 (2003).


\bibitem{Schirmacher_VH}
W. Schirmacher, G. Diezemann, and C. Ganter, Phys. Rev. Lett. 81, 136 (1998).

\bibitem{Zorn}
R. Zorn, Phys. Rev. B 81, 054208 (2010).

\bibitem{Schirmacher}
W. Schirmacher, G. Ruocco, and T. Scopigno, Phys. Rev. Lett. 98, 025501 (2007); W. Schirmacher, Europhys. Lett. 73, 892 (2006); W. Schirmacher, J. Non-Cryst. Solids 357, 518 (2011).

\bibitem{Schirmacher2}
W. Schirmacher, Phys. Stat. Sol. B 250, 937 (2013).





\bibitem{Binder}
K. Binder and D. Heermann, \textit{Monte-Carlo simulation in Statistical Physics} (Springer-Verlag, Berlin Heidelberg, 2010).

\bibitem{Thorpe}
M.F. Thorpe, J. Non-Cryst. Solids \textbf{57}, 355 (1983).

\bibitem{Ashcroft}
Ashcroft, N.W. and Mermin, N. D. \textit{Solid State Physics} (Thomson Brooks/Cole,
1976).


\bibitem{Lemaitre}
A. Lemaitre and C. Maloney,  J. Stat. Phys. \textbf{123}, 415 (2006).
	
\bibitem{Cady}
W. G. Cady, \textit{Piezoelectricity} (Dover, New York, 1962). 

\bibitem{Zaccone2011}
A. Zaccone and E. Scossa-Romano,  Phys. Rev. B {\bf 83}, 184205 (2011); A. Zaccone, J. Blundell, and E. M. Terentjev, Phys. Rev. B {\bf 84}, 174119 (2011).

\bibitem{Yoshino}
H. Yoshino,  J. Chem. Phys. \textbf{136}, 214108 (2012).

\bibitem{Vitelli2}
N. Xu, V. Vitelli, A.J. Liu, S.R. Nagel, EPL \textbf{90}, 56001 (2010). 

\bibitem{Frenkel}
S. Auer and D. Frenkel, J. Chem. Phys. 120, 3015 (2004); J. Russo and H. Tanaka, Sci. Rep. \textbf{2}, 505 (2012).


\bibitem{Bonjour}
E. Bonjour, R. Calemczuk, R. Lagnier, E. Salce, J. de Phys. \textbf{42}, 63 (1981).


\bibitem{Ramos} A.M. Ramos et al.,  Phys. Rev. Lett. \textbf{78}, 82 (1997).

\bibitem{Zamponi} 
S. Franz, G. Parisi, P. Urbani, F. Zamponi,  Proc. Natl. Acad. Sci. USA \textbf{112}, 14539 (2015).

\bibitem{Wyart}
M. Wyart, EPL 89, 64001 (2010). 

\bibitem{Barrat}
F. Leonforte, A. Tanguy, J.P. Wittmer, and J.-L. Barrat, Phys. Rev. Lett. \textbf{97}, 055501 (2006).

\bibitem{Mizuno}
H. Mizuno, S. Mossa, J.-L. Barrat,  Proc. Natl. Acad. Sci. USA 111, 11949 (2014). 

\bibitem{GarrahanPRL}
J. P. Garrahan and D. Chandler,  Phys. Rev. Lett. \textbf{89}, 035704 (2002). 

\bibitem{Garrahan1} 
R. L. Jack, L. O. Hedges, J. P. Garrahan, and D. Chandler,  Phys. Rev. Lett. \textit{107}, 275702 (2011).

\bibitem{Garrahan2} 
D.J. Ashton and J.P. Garrahan,  Eur. Phys. J. E \textbf{30}, 303–307 (2009)

 














%
%
%
%




\end{thebibliography}
\end{document}